\documentclass[aps,pra,superscriptaddress,twocolumn,notitlepage,longbibliography]{revtex4-1}
\usepackage{amsmath}
\usepackage{graphicx}
\usepackage{rotating}
\usepackage{amssymb}
\usepackage{txfonts} 
\usepackage[usenames,dvipsnames]{color}

\usepackage[colorlinks=true,
            linkcolor=Blue,
            citecolor=Blue,
            urlcolor=Blue,
            pdftitle={Rydberg Fermi Gas},
            pdfauthor={A. Keles},
            bookmarks=true
            ]{hyperref}

\usepackage{float}

\usepackage[numbered,open]{bookmark}

\usepackage[all]{hypcap}

\graphicspath{ {./figs/} }

\usepackage{listings}

\usepackage{tikz}

\tikzset{middlearrow/.style={
    decoration={markings,
      mark= at position 0.55 with {\arrow[scale=1,blue]{#1}} ,
    },
    postaction={decorate}
  }
}
\usetikzlibrary{decorations.pathreplacing,
  decorations.markings,decorations.pathmorphing,
    calc,
  shapes}

\usepackage{multirow}

\usepackage{pifont}

\newcommand\x{\ensuremath\mathbf{x}}

\newcommand\qb{\ensuremath\mathbf{q}}
\newcommand\ef{\ensuremath\epsilon_F}

\usepackage{bbold}

\newcommand\w{\omega}

\newcommand\pb{\mathbf{p}}

\newcommand{\oneloop}[1]
{
\begin{tikzpicture}[scale=.4,baseline=(current bounding box.center)]

    \def \x {0}
    \def \y {.1}
    \def \w {.3}
    \def \l {.5}  
    \def \dver {3}
    \draw[fill=gray] (\x,\y) rectangle (\x+\w,\y+\w);
    \draw[fill=gray] (\x+\w+\dver,\y) rectangle (\x+\dver+2*\w,\y+\w);
    \foreach \m/\n [count=\i] in {#1}
    {
      \ifthenelse{\equal{\i}{1}}
      {
        \ifx\m\n
          \draw[middlearrow={stealth}] 
        \else
          \draw[middlearrow={stealth reversed}] 
        \fi
        (\x-\l,\y+\w+\l) -- (\x,\y+\w);
        \node [left, black] at (\x-\l,\y+\w+\l) 
        {\tiny\ensuremath{\m}};
      }{}  
      \ifthenelse{\equal{\i}{2}}
      {
        \ifx\m\n
        \draw[middlearrow={stealth}] 
        \else
        \draw[middlearrow={stealth reversed}] 
        \fi
        (\x-\l,\y-\l) -- (\x,\y);
        \node [left, black] at (\x-\l,\y-\l) 
        {\tiny\ensuremath{\m}};
      }{} 
      \ifthenelse{\equal{\i}{3}}
      {
        \ifx\m\n
        \draw[middlearrow={stealth}] 
        \else
        \draw[middlearrow={stealth reversed}] 
        \fi
        (\x+2*\w+\dver,\y+\w) -- (\x+2*\w+\dver+\l,\y+\w+\l);
        \node [right, black] at (\x+2*\w+\dver+\l,\y+\w+\l) 
        {\tiny\ensuremath{\m}};
      }{} 
      \ifthenelse{\equal{\i}{4}}
      {
        \ifx\m\n
        \draw[middlearrow={stealth}] 
        \else
        \draw[middlearrow={stealth reversed}] 
        \fi
        (\x+2*\w+\dver,\y) -- (\x+2*\w+\dver+\l,\y-\l);
        \node [right, black] at (\x+2*\w+\dver+\l,\y-\l) 
        {\tiny\ensuremath{\m}};
      }{} 
      \ifthenelse{\equal{\i}{5}}
      {
        \ifx\m\n
        \draw[middlearrow={stealth reversed}] 
        \else
        \draw[middlearrow={stealth }] 
        \fi
        (\x+\w+\dver,\y+\w) 
        to [out=145,in=35]
        (\x+\w,\y+\w);
        \node [above, black] at (\x+\w+.5*\dver,\y+\w+.5)
        {\tiny\ensuremath{\m}};
      }{} 
      \ifthenelse{\equal{\i}{6}}
      {
        \ifx\m\n
        \draw[middlearrow={stealth reversed}] 
        \else
        \draw[middlearrow={stealth }] 
        \fi
        (\x+\w+\dver,\y) 
        to [out=-145,in=-35] 
        (\x+\w,\y);
        \node [below, black] at (\x+\w+.5*\dver,\y-0.5) 
        {\tiny\ensuremath{\m}};
      }{} 
    } 
  \end{tikzpicture}
}

\def\iline(#1,#2,#3,#4,#5,#6)
{
  \draw[
  thick,
  decorate,
  decoration={snake, amplitude=0.3mm, segment length=1.0mm}
  ]
  (#1,#2) -- (#3,#4); 
}
\def\fline(#1,#2,#3,#4,#5,#6)
{
  \draw[middlearrow={stealth}] (#1,#2) to (#3,#4); 
} 
\newcommand{\vertex}[1]
{
\begin{tikzpicture}[scale=.3,baseline=(current bounding box.center)]

    \def \x {0}
    \def \y {.1}
    \def \w {.5}
    \def \l {.6}  
    \def \dver {3}
    \draw[fill=gray] (\x,\y) rectangle (\x+\w,\y+\w);
    \foreach \m/\n [count=\i] in {#1}
    {
      \ifthenelse{\equal{\i}{1}}
      {
        \ifx\m\n
          \draw[middlearrow={stealth}] 
        \else
          \draw[middlearrow={stealth reversed}] 
        \fi
        (\x-\l,\y+\w+\l) -- (\x,\y+\w);
        \node [left, black] at (\x-\l,\y+\w+\l) 
        {\tiny\ensuremath{\m}};
      }{}  
      \ifthenelse{\equal{\i}{2}}
      {
        \ifx\m\n
        \draw[middlearrow={stealth}] 
        \else
        \draw[middlearrow={stealth reversed}] 
        \fi
        (\x-\l,\y-\l) -- (\x,\y);
        \node [left, black] at (\x-\l,\y-\l) 
        {\tiny\ensuremath{\m}};
      }{} 
      \ifthenelse{\equal{\i}{3}}
      {
        \ifx\m\n
        \draw[middlearrow={stealth}] 
        \else
        \draw[middlearrow={stealth reversed}] 
        \fi
        (\x+\w,\y+\w) -- 
        (\x+\w+\l,\y+\w+\l);
        \node [right, black] at (\x+\w+\l,\y+\w+\l)
        {\tiny\ensuremath{\m}};
      }{} 
      \ifthenelse{\equal{\i}{4}}
      {
        \ifx\m\n
        \draw[middlearrow={stealth }] 
        \else
        \draw[middlearrow={stealth reversed}] 
        \fi
        (\x+\w,\y) -- 
        (\x+\w+\l,\y-\l);
        \node [right, black] at (\x+\w+\l,\y-\l) 
        {\tiny\ensuremath{\m}};
      }{} 
    } 
  \end{tikzpicture}
}
\newcommand{\oneloopV}[1]
{
  \begin{tikzpicture}[scale=.8, baseline=(current bounding box.center)]
    \ifthenelse{\equal{1}{#1}}
    {
      \iline(1,0,1,1,d,a )  
      \iline(2,0,2,1,d,a )  

      \fline(0.5,0,1,0,d,a )  
      \fline(1,0,2,0,d,a )  
      \fline(2,0,2.5,0,d,a )  

      \fline(0.5,1,1,1,d,a )  
      \fline(1,1,2,1,d,a )  
      \fline(2,1,2.5,1,d,a )
    }{}  
    \ifthenelse{\equal{2}{#1}}
    {
      \iline(1.5, 0.0, 1.5, 0.3, d, a )  
      \iline(1.5, 0.7, 1.5, 1.0, d, a )  

      \fline(0.5, 0.0, 1.5, 0.0, d, a )  
      \fline(1.5, 0.0, 2.5, 0.0, d, a )  

      \fline(0.5, 1.0, 1.5, 1.0, d, a )  
      \fline(1.5, 1.0, 2.5, 1.0, d, a )  
      \draw[middlearrow={stealth}] (1.5, 0.7) to[out=0,in=0, looseness=1.5]  (1.5,0.3); 
      \draw[middlearrow={stealth}] (1.5, 0.3) to[out=180,in=180, looseness=1.5]  (1.5,0.7);
    }{}  
    \ifthenelse{\equal{3}{#1}}
    {
      \fline(1,0,1.5,.5,d,a )  
      \fline(1.5,.5,2,0,d,a )  
      \iline(1.5,.5,1.5,1,d,a )  

      \fline(0.5,0,1,0,d,a )  
      \iline(1,0,2,0,d,a )  
      \fline(2,0,2.5,0,d,a )  

      \fline(0.5, 1.0, 1.5, 1.0, d, a )  
      \fline(1.5, 1.0, 2.5, 1.0, d, a )
    }{}  
    \ifthenelse{\equal{4}{#1}}
    {
      \fline(1.0, 1.0, 1.5, 0.5, d, a )  
      \fline(1.5, 0.5, 2.0, 1.0, d, a )  
      \iline(1.5, 0.0, 1.5, 0.5, d, a )  

      \iline(1.0, 1.0, 2.0, 1.0, d, a )  
      \fline(0.5, 0.0, 1.5, 0.0, d, a )  
      \fline(1.5, 0.0, 2.5, 0.0, d, a )  

      \fline(0.5, 1.0, 1.0, 1.0, d, a )  
      \fline(2.0, 1.0, 2.5, 1.0, d, a )
    }{}  
    \ifthenelse{\equal{5}{#1}}
    {
      \iline(1,0,2,1,d,a )  
      \iline(2,0,1,1,d,a )  

      \fline(0.5,0,1,0,d,a )  
      \fline(1,0,2,0,d,a )  
      \fline(2,0,2.5,0,d,a )  

      \fline(0.5,1,1,1,d,a )  
      \fline(1,1,2,1,d,a )  
      \fline(2,1,2.5,1,d,a )
    }{}  

  \end{tikzpicture}
}

\newcommand{\siteVertexA}[1]
{
  \begin{tikzpicture}[scale=.8, baseline=(current bounding box.center)]
      \iline(1,0.25,1,.75,d,a )  

      \fline(0.5,0,1,0.25,d,a )  
      \fline(1,0.25,1.5,0,d,a )  

      \fline(0.5,1,1,.75,d,a )  
      \fline(1,0.75,1.5,1,d,a )  
    \node [above, black] at (1,.8) {\tiny\ensuremath{i_1}};
    \node [below, black] at (1,.2) {\tiny\ensuremath{i_2}};
    \node [left, black] at (0.5,1) {\tiny\ensuremath{1}};
    \node [left, black] at (0.5,0) {\tiny\ensuremath{2}};
    \node [right, black] at (1.5,1){\tiny\ensuremath{1'}};
    \node [right, black] at (1.5,0){\tiny\ensuremath{2'}};
  \end{tikzpicture}
}

\newcommand{\siteVertexB}[1]
{
  \begin{tikzpicture}[scale=.8, baseline=(current bounding box.center)]
      \iline(1,0.25,1,.75,d,a )  

      \fline(0.5,0,1,0.25,d,a )  
      \fline(1,0.25,1.5,0,d,a )  

      \fline(0.5,1,1,.75,d,a )  
      \fline(1,0.75,1.5,1,d,a )  
    \node [above, black] at (1,.8) {\tiny\ensuremath{i_1}};
    \node [below, black] at (1,.2) {\tiny\ensuremath{i_2}};
    \node [left, black] at (0.5,1) {\tiny\ensuremath{1}};
    \node [left, black] at (0.5,0) {\tiny\ensuremath{2}};
    \node [right, black] at (1.5,1){\tiny\ensuremath{2'}};
    \node [right, black] at (1.5,0){\tiny\ensuremath{1'}};
  \end{tikzpicture}
}

\begin{document}
\title{$f$-wave superfluidity from repulsive interaction in Rydberg-dressed Fermi gas }
\author{Ahmet Kele\c{s}} 
\affiliation{Department of Physics and Astronomy,
  University of Pittsburgh, Pittsburgh, Pennsylvania 15260, USA}
\affiliation{Department of Physics and Astronomy,
  George Mason University, Fairfax, Virginia 22030, USA}
\author{Erhai Zhao} 
\affiliation{Department of Physics and Astronomy, George Mason University, Fairfax, Virginia 22030,
  USA}
\author{Xiaopeng Li} 
\email{xiaopeng\_li@fudan.edu.cn}
\affiliation{
State Key Laboratory of Surface Physics, Institute of Nanoelectronics and 
Quantum Computing, \& Department of Physics, Fudan University, Shanghai 200433, China}
\begin{abstract}

Interacting Fermi gas provides an ideal model system to understand unconventional pairing 
and intertwined orders relevant to a large class of quantum materials.  
 Rydberg-dressed Fermi gas is a recent experimental
 system where the sign, strength, and range of the interaction can be
 controlled.  The interaction in momentum space has a negative
 minimum at $q_c$ inversely proportional to the characteristic length-scale in
 real space, the soft-core radius $r_c$. We show theoretically that
 single-component  (spinless) Rydberg-dressed Fermi gas in two dimensions has
 a rich phase diagram with novel superfluid and density wave orders due to the
 interplay of the Fermi momentum $p_F$, interaction range $r_c$, and
 interaction strength $u_0$. For repulsive bare interactions $u_0>0$, the
 dominant instability is $f$-wave superfluid for $p_Fr_c\lesssim 2$,
 and density wave for $p_Fr_c\gtrsim 4$. The $f$-wave pairing in this repulsive
 Fermi gas is reminiscent of the conventional
 Kohn-Luttinger mechanism, but has a much higher $T_c$.  For attractive bare interactions $u_0<0$, the
 leading instability is $p$-wave pairing. 
 The phase diagram is obtained from
 functional renormalization group 
 that treats all
 competing many-body instabilities in the particle-particle and particle-hole
 channels on equal footing.

\end{abstract}

 \maketitle

\section{Introduction} 

Understanding the many-body instabilities and symmetry breaking
in strongly interacting fermions in two-dimension (2D) holds the key to
several long-standing
problems in condensed matter physics.
 One example is the precise mechanism by which unconventional
superconductivity with various pairing symmetries emerges from repulsive
interactions, 
in materials ranging from cuprate \cite{RevModPhys.78.17}, ruthenate
\cite{ruthenates}, and pnictide \cite{iron} 
superconductors. 
These and other correlated quantum materials typically display
intertwined vestigial orders, e.g. in the so-called pseudogap 
region  where charge density waves, pairing, and other fluctuations compete.
Recently, ultracold Fermi
gases \cite{RevModPhys.80.1215,ketterle2008making} of atoms and molecules have become a promising experimental platform to
tackle some of these open 
problems by realizing 
Hamiltonians such as the  Fermi-Hubbard  model \cite{hart2015observation,mazurenko2017cold,brown2019bad} 
with tunable interactions \cite{RevModPhys.80.885}. 
%
This offers  opportunity to deepen our understanding of the ``pairing glue" in repulsively 
interacting systems, and shed light on the complex interplay of quantum fluctuations 
in distinct channels for simple and highly controlled 
Hamiltonians.
In this paper, we show theoretically that Rydberg-dressed Fermi gas of alkali
atoms with tunable long-range interactions gives rise
to not only $p$-wave topological superfluids for attractive bare interactions,
but also $f$-wave superfluid with high transition temperatures stemming from repulsive bare interactions. 

Rydberg atoms and Rydberg-dressed atoms haven long been recognized for their
potential in quantum simulation and quantum information \cite{Weimer2010,
  PhysRevLett.87.037901,Saffman2010,Browaeys2016,Karpiuk_2015}.
%
Recent experiments have successfully demonstrated a panoply of two-body
interactions in cold gases of Rydberg-dressed alkali atoms
\cite{schauss2015crystallization,zeiher2016many,holletith2018,
  PhysRevX.7.041063,Jau2015,PhysRevX.8.021069,borish2019transversefield}.
In Rydberg dressing, the
ground state atom (say $n_0S$) is weakly coupled to a Rydberg state (say $nS$
or $nD$) with large principal number $n$ by off-resonant light with Rabi
frequency $\Omega$ and detuning $\Delta$. The coupling can be achieved for
example via a two-photon process involving an intermediate state $n_1 P$ to
yield longer coherence times \cite{Henkel2010}. The huge
dipole moments of the Rydberg states lead to strong interactions that 
exceed the natural van der Waals interaction by a factor that scales with
powers
of $n$ \cite{Saffman2010,Browaeys2016}.  
The
interaction between two Rydberg-dressed atoms takes the following form
\cite{Henkel2010}:
\begin{equation}
  V(\mathbf{r}) = \frac{u_0}{r^6+r_c^6}.
  \label{eq:rydberg_interaction_real}
\end{equation}
Here $r=|\mathbf{r}|$ is the inter-particle distance,
$u_0=(\Omega/2\Delta)^4C_6$ is the interaction strength, $C_6$ is the
van der Waals coefficient, and $r_c=|C_6/2\hbar\Delta|^{1/6}$ is 
the soft-core radius and the characteristic scale for the interaction
range.  As shown in Fig.~\ref{fig:rydberg_interaction}, $V({\bf r})$ has a
step-like soft-core for $r\lesssim r_c$ before decaying to a van der Waals
tail at long distances.
Both $u_0$ and $r_c$ can be tuned experimentally
via $\Omega$ and $\Delta$ \cite{Henkel2010}.  Moreover, by choosing proper
Rydberg states (e.g. $nS$ versus $nD$ for $^6$Li with $n>30$ \cite{Xiong2014})
$C_6$ and $u_0$ can be made either repulsive or attractive. 
By choosing proper $n$, $\Delta$ and $\Omega$, atom loss can be reduced to achieve a 
sufficiently long life time to observe many-body phenomena
\cite{Henkel2010, PhysRevX.7.041063, PhysRevX.8.021069,Li2016a}.

Previous theoretical studies have explored the novel many-body phenomena
associated with interaction Eq.~\eqref{eq:rydberg_interaction_real} in 
bosonic
\cite{Henkel2010,Maucher2011,Glaetzle2014,tanatar2019,Pupillo2010,PhysRevLett.108.265301,PhysRevLett.123.045301,PhysRevLett.119.215301} 
and fermionic gases \cite{Li2016a} including the prediction of topological
superfluids \cite{Xiong2014} and topological density waves \cite{Li2015a}.
Here we consider single-component Rydberg Fermi gases confined in 2D
\cite{Khasseh2017}, where mean-field and random phase approximation (RPA)
become unreliable due to enhanced quantum fluctuations. Our goal is to set
up a theory to systematically describe the competing many-body phases of 2D
Rydberg-dressed Fermi gas by treating them on equal footing beyond the
weak-coupling regime and RPA. 
We achieve this by solving the functional renormalization group flow equations
for the fermionic interaction vertices. The resulting phase diagram
(Fig.~\ref{fig:phase_diagram}) 
is much richer than the RPA prediction \cite{Khasseh2017} and reveals 
an unexpected $f$-wave phase.

The paper is organized as follows. In Sec.~\ref{sec:mean-field} we introduce
many-body phases of Rydberg-dressed Fermi gas within mean-field from the
standard Cooper
instability analysis and Random Phase Approximation. In Sec.~\ref{sec:frg} we
present the numerical implementation of Functional Renormalization
Group to this problem and in Sec.~\ref{sec:results} we show many-body phases
beyond mean field calculation which manifest intertwined quantum
fluctuations in pairing and density-wave channels. In
Sec.~\ref{sec:conclusions}, we summarize our study and implications of our
findings for
future experimental developments in ultracold gases. 

\section{Rydberg-dressed Fermi gas}
\label{sec:mean-field}
\begin{figure}
  \centering
  \includegraphics[scale=.5]{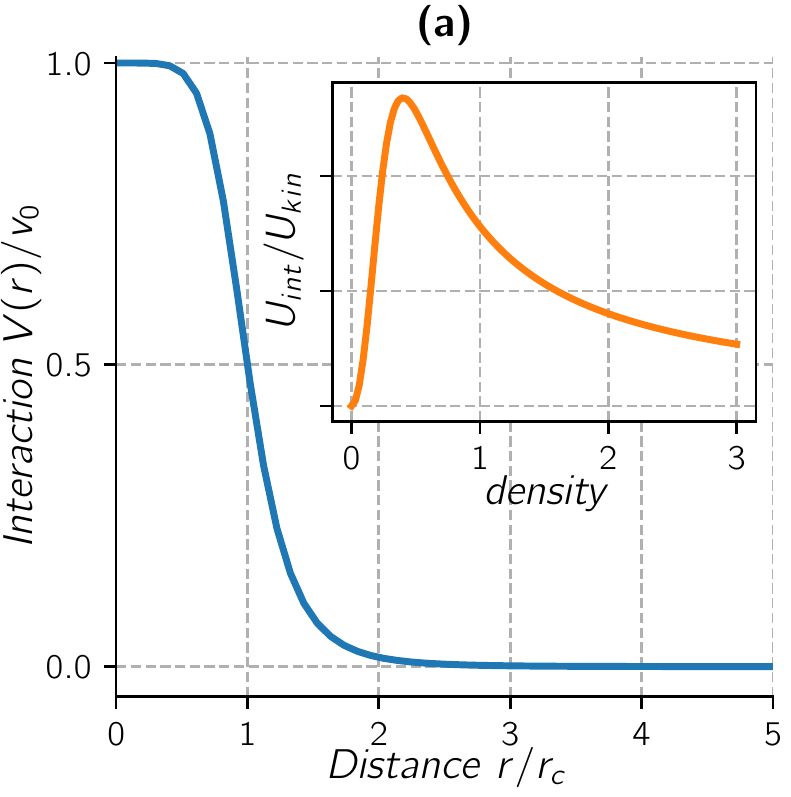}
  \includegraphics[scale=.5]{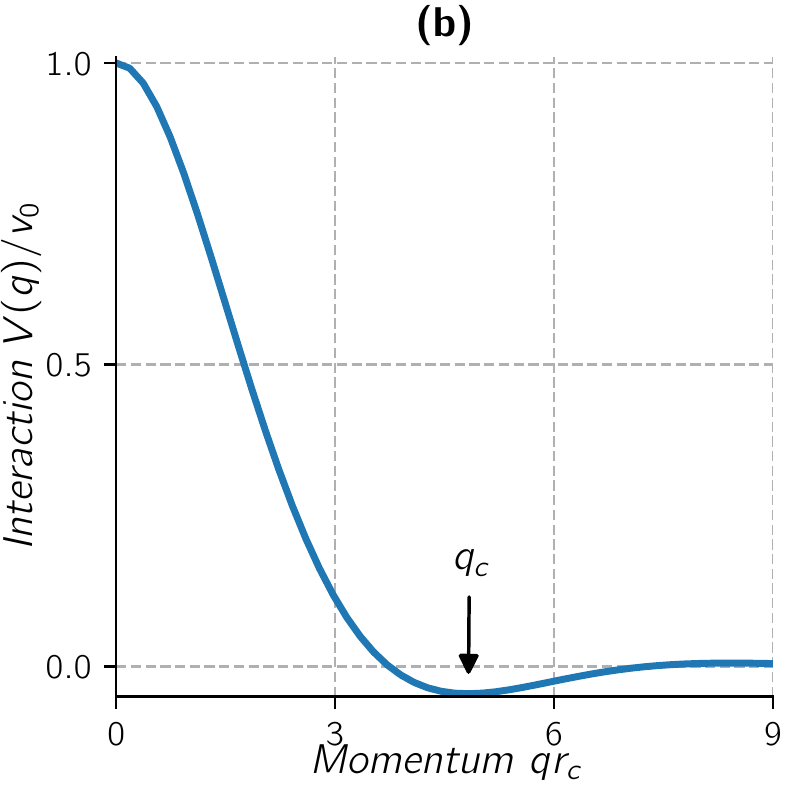}
  \includegraphics[scale=.5]{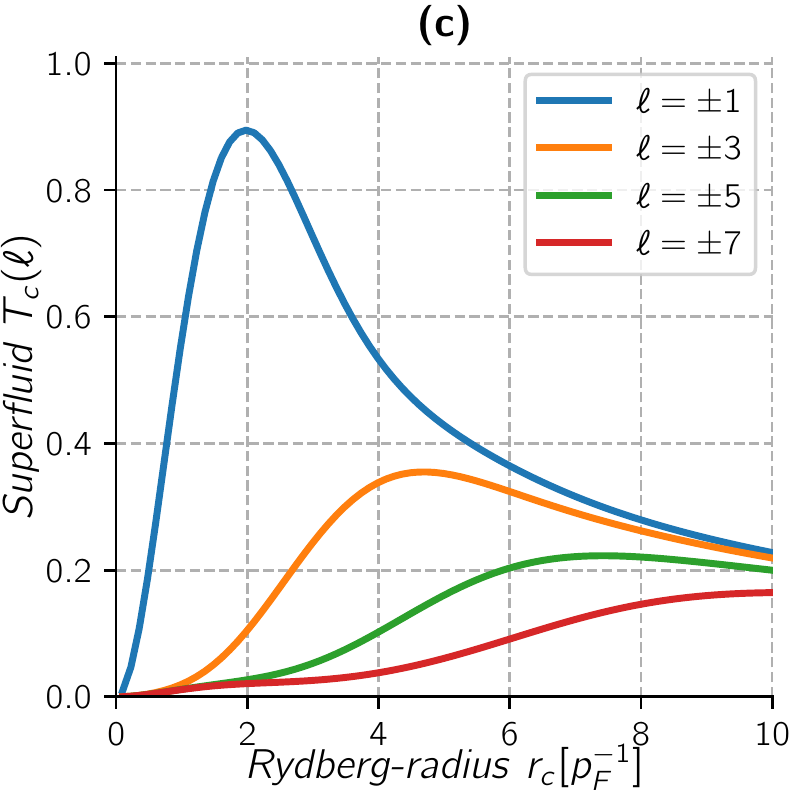}
  \includegraphics[scale=.5]{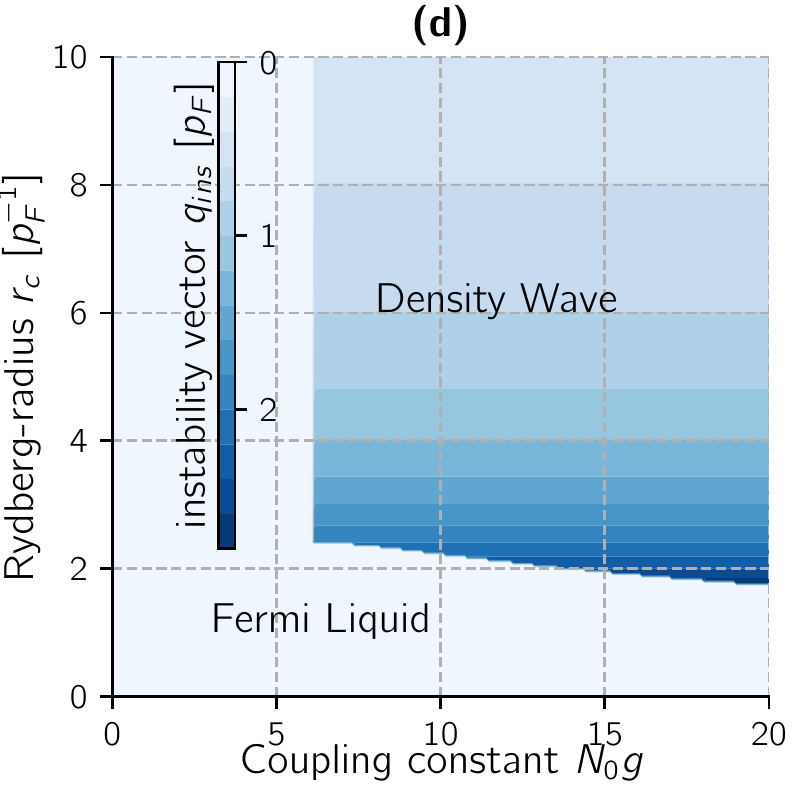}
  \caption{Single-component Fermi gas with Rydberg-dressed interactions in 2D.
    (a) The interaction potential Eq. \eqref{eq:rydberg_interaction_real}
    shows a step-like soft core of radius $r_c$ and a long-range tail.
    (Inset) Ratio of the interaction to kinetic energy 
    varies non-monotonically with density.  (b) The Rydberg-dressed
    interaction Eq. \eqref{eq:rydberg_interaction_momentum} in momentum space
    attains a negative minimum at $q_c\sim 4.82/r_c$.  (c) For attractive
    interactions, the critical temperatures in different angular momentum
    $\ell$ channels (in arbitrary units) from the solution of the Cooper
    problem. The leading instability is $p$-wave, $\ell=\pm 1$.  Maximum $T_c$
    is around $p_Fr_c\approx 2$. (d) For repulsive interactions, random phase
    approximation points to a density-wave order. False color (shading) shows the
    ordering wave vector of density modulations.}
  \label{fig:rydberg_interaction}
\end{figure}

We first highlight the unique properties of
Rydberg-dressed Fermi gas by comparing it with other well-known Fermi systems
with long-range interactions such as the electron gas and dipolar Fermi gas.
Correlations in electron liquid are characterized by a single
dimensionless parameter $r_s$, the ratio of Coulomb interaction energy to
kinetic energy. In the high density limit $r_s\ll 1$, the system
is weakly interacting while in the low density limit $r_s\gg 1$, Wigner
crystal is formed. The intermediate correlated regime with $r_s\sim 1$ can
only be described by various approximations \cite{nozieres-pines}. Similarly,
dipolar Fermi gas also has a power-law interaction that lacks a scale, so a
parameter analogous to $r_s$ can be introduced which varies monotonically with
the density \cite{Baranov2012a}.
The situation is different in Rydberg-dressed Fermi gas with interaction given
by Eq.~\eqref{eq:rydberg_interaction_real}. From the inter-particle spacing
$1/\sqrt{2\pi n}$ and the Fermi energy $\ef=2\pi n/m$ (we put $\hbar=1$ and
$k_B=1$) in terms of areal density $n$, one finds that the ratio of interaction
energy to kinetic energy scales as $n^2/[1+(2\pi r_c^2)^3 n^3]$, which varies
non-monotonically with $n$ unlike electron liquid due to $r_c$
(Fig.~\ref{fig:rydberg_interaction}a inset).  
Distinctive feature of the interaction $V(\mathbf{r})$ is revealed by its
Fourier transform in 2D \cite{Khasseh2017},
\begin{equation}
  V(\qb) = g 
  G\left(
     {q^6r_c^6}/{6^6}
 \right),\;\;\; g=\pi u_0/3r_c^4,
 \label{eq:rydberg_interaction_momentum}
\end{equation}
where $\mathbf{q}$ is the momentum, $q=|\mathbf{q}|$, $g$ is the coupling strength
and
$G$ is the Meijer G-function \footnote{
  In Mathematica, this Meijer G-function is called with
  \texttt{
    MeijerG\big[\{\{\},\{\}\},\{\{0,1/3,2/3,2/3\},\{0,1/3\}\},$z^6/6^6$\big]
  } where $z=qr_c$.
}.
The function $V(\qb)$, plotted in Fig.~\ref{fig:rydberg_interaction}b,
develops a negative minimum at $q=q_c\sim 4.82/r_c$. This is the momentum
space manifestation of the step-like interaction potential
Eq.~\eqref{eq:rydberg_interaction_real}.  These unique behaviors 
are the main culprits of its rich 
phase diagram.

Starting from the free Fermi gas, 
increasing the interaction $g$ may lead to a diverging susceptibility and 
drive the Fermi liquid into a symmetry-broken 
phase.  
We first give a qualitative discussion of potential ordered phases
using standard methods to orient our numerical FRG results later.
For attractive interactions, $u_0<0$, an arbitrarily small $g$ is sufficient
to drive the Cooper instability.  By decomposing $V(\qb=\pb_F-\pb_F')$ into
angular momentum channels, $V(2p_F\sin\frac{\theta}{2})=\sum_\ell V_\ell
e^{i\ell \theta}$ where $\theta$ is the angle between $\pb_{F}$ and $\pb_F'$,
one finds different channels decouple and the critical
temperature of the $\ell$-th channel $T_c(\ell) \sim e^{-1/N_0V_\ell}$
\cite{mineev1999} with 
$N_0=m/2\pi$ being the density of states.
Thus the
leading instability comes from the channel with the largest $V_\ell$ (hence
the largest $T_c$). Fig.~\ref{fig:rydberg_interaction}c illustrates
$T_c(\ell)$ as a function of $r_c$ for fixed $p_F$. It is apparent that the
dominant instability is in the $\ell=\pm 1$ channel, i.e., $p$-wave pairing.
Its $T_c$ develops a dome structure and reaches maximum around $p_Fr_c\approx
2$.  For large $r_c$, higher angular momentum channels start to compete with
the $\ell=\pm 1$ channel. 

For repulsive bare interactions, $u_0>0$, a sufficiently strong interaction
$g$ can induce an instability toward the formation of (charge) density waves.
This has been shown recently \cite{Khasseh2017} for 2D Rydberg-dressed Fermi
gas using random phase approximation (RPA) which sums over a geometric series
of ``bubble diagrams" to yield the static dielectric function,
$\epsilon(\qb)=1-V(\qb)\chi_0(\qb)$ where the Linhard function
$\chi_0(\qb)=-N_0[ 1- \Theta(q-2k_F) \sqrt{q^2-4k_F^2}/q]$. The onset of
density wave instability is signaled by $\epsilon(\qb)=0$ at some wave vector
$q=q_{ins}$, i.e. the softening of particle-hole excitations.  Within RPA,
$q_{ins}$ always coincides with $q_c$, and the resulting phase diagram is
shown in Fig.~\ref{fig:rydberg_interaction}d.

While these standard considerations capture the $p$-wave pairing and density
wave order, they fail to describe the physics of
intertwined scattering between particle-particle and particle-hole channels.
We show below that this missing ingredient exhibits significant effects,
leading to the emergence of a robust $f$-wave superfluid in the repulsive
regime. 
For a detailed comparison between RPA and FRG see Ref.
\cite{PhysRevA.94.033616}.

\section{Numerical Implementation of Functional Renormalization Group}
\label{sec:frg}
Functional renormalization group (FRG) is a powerful
technique that can accurately predict the many-body instabilities of strongly
interacting fermions \cite{RevModPhys.84.299}. It implements Wilson's
renormalization group for interacting fermions in a formally exact manner by
flowing the generating functional of the many-body system $\Gamma$ as a sliding
momentum scale $\Lambda$ is varied. Starting from the bare
interaction $V(\qb)$ at a chosen ultraviolet scale $\Lambda_{UV}$, higher
energy fluctuations are successively integrated out to yield the self-energy
$\Sigma$ and effective interaction vertex $\Gamma$ at a lower scale
$\Lambda<\Lambda_{UV}$. As $\Lambda$ is lowered toward a very small value
$\Lambda_{IR}$, divergences in the channel coupling matrices and
susceptibilities point to the development of long-range order.  Its advantage
is that all ordering tendencies are treated unbiasedly with full momentum
resolution. The main draw back is its numerical complexity:  at each RG step,
millions of running couplings have to be retained. 
FRG has been applied to dipolar Fermi gas
\cite{PhysRevA.94.033616,PhysRevLett.108.145301} and extensively benchmarked
against different techniques
\cite{PhysRevA.84.063633, PhysRevLett.108.145304, PhysRevB.84.235124,
  PhysRevA.84.063633, PhysRevB.91.224504}.
For more details about the formalism, see reviews \cite{RevModPhys.84.299} and
\cite{peterkopietz2010}. Note that our system is a continuum Fermi gas, not a
lattice system extensively studied and reviewed in \cite{RevModPhys.84.299}.

The central task of FRG is to solve the coupled flow equations for self-energy
$\Sigma_{1',1}$ and two-particle vertex $\Gamma_{1',2';1,2}$
\cite{RevModPhys.84.299}:
\begin{align}
    \partial_\Lambda \Sigma_{1',1} &= -\sum_{2} S_{2} \Gamma_{1',2;1,2},
    \nonumber\\
    \partial_\Lambda\Gamma_{1',2';1,2} &= \sum_{3,4} \Pi_{3,4} 
    \big[
      \frac{1}{2} \Gamma_{1',2';3,4} \Gamma_{3,4;1,2} 
      -\Gamma_{1,'4;1,3}\Gamma_{3,2';4,2} 
      \nonumber\\
     &+\Gamma_{2',4;1,3}\Gamma_{3,1';4,2} 
    \big],
    \label{eq:flow}
\end{align}
Here the short-hand notation $1\equiv(\w_1,\pb_1)$, $1,2$ ($1',2'$)
label the incoming (outgoing) legs of the four-fermion vertex $\Gamma$, and
the sum stands for integration over frequency 
and momentum, $\Sigma\rightarrow \int d\w d^2\pb/(2\pi)^3$. 
Diagrammatically, the first term in Eq. \eqref{eq:flow} is the BCS diagram in
the particle-particle channel, and the second and third terms are known as the
ZS and ZS' diagram in the particle-hole channel \cite{RevModPhys.66.129}.  
The polarization bubble  $\Pi_{3,4} = G_{3} S_{4} + S_{3} G_{4}$ 
contains the product of two scale-dependent Green functions defined by
\begin{align}
    G_{\w,\pb} =
    \frac{\Theta(|\xi_\pb|-\Lambda) }{i\w-\xi_\pb-\Sigma_{\w,\pb}  } ,\quad
    \quad
    S_{\w,\pb} =
    \frac{\delta(|\xi_\pb|-\Lambda) }{i\w-\xi_\pb-\Sigma_{\w,\pb}  }.
    \label{eq:propagators}
\end{align}
Note that $G$, $S$, $\Sigma$ and $\Gamma$ all depend on the sliding scale
$\Lambda$, we suppressed their $\Lambda$-dependence in equations above for
brevity.

\begin{figure}
  \centering
  \includegraphics[width=0.4\textwidth]{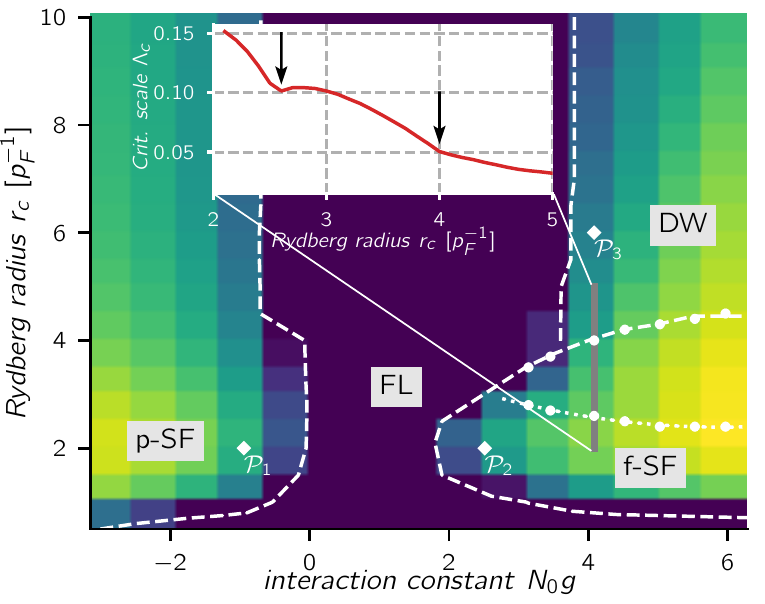}
  \includegraphics[width=0.4\textwidth]{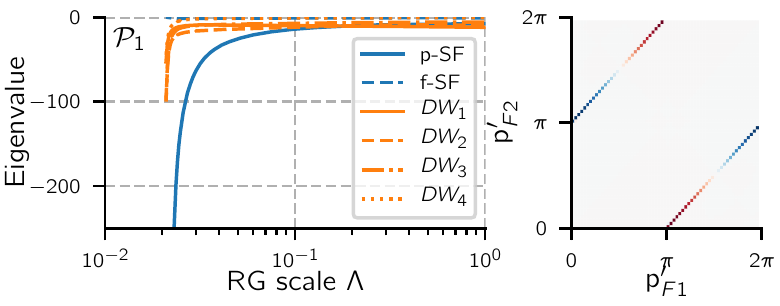}
  \includegraphics[width=0.4\textwidth]{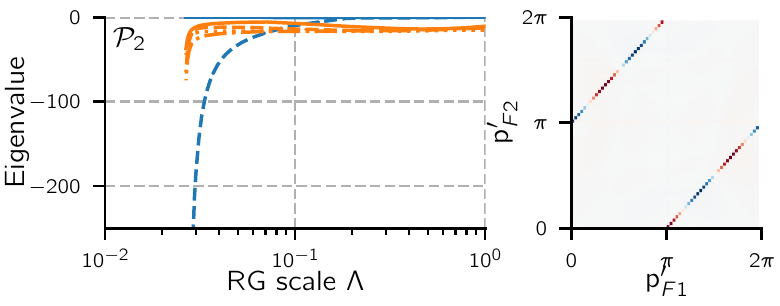}
  \includegraphics[width=0.4\textwidth]{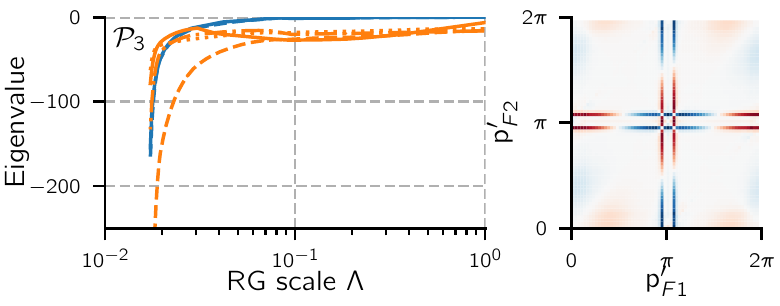}
  \caption{Phase diagram of Rydberg-dressed spinless Fermi gas in 2D based on
    FRG.  Tuning the interaction range $r_c$ and interaction strength $g$
    yields Fermi Liquid (FL), $p$-wave superfluid (p-SF), $f$-wave superfluid
    (f-SF), and density-wave (DW).  False color (shading) indicates the
    critical scale
    $\Lambda_c$ of the instability where brighter (darker) regions have
    higher (lower) $T_c$.  Panels labelled with $\mathcal{P}_1$,
    $\mathcal{P}_2$ and $\mathcal{P}_3$ show the details of renormalization
    flow and vertex function for points marked with white diamonds on the
    phase diagram.  The leading eigenvalues for a few channels (see legends)
    are shown on the left.  The maps of vertex function
    $\Gamma(\pb_{F1}',\pb_{F2}',\pb_{F1})$ are shown on the right for fixed
    $\pb_{F1}=(-p_F,0)$.  Superfluid (density wave) order displays diagonal
    (horizontal and vertical) correlations.}
  \label{fig:phase_diagram}
\end{figure}

Several well-justified approximations are used to make the flow equations
computationally tractable.  To identify leading
instabilities, the self-energy can be safely dropped, and the frequency
dependence of $\Gamma$ can be neglected \cite{RevModPhys.84.299}.  As a
result, the frequency integral of the fermion loops in Eq. \eqref{eq:flow} can
be performed analytically. Furthermore, we retain the most relevant dependence of
$\Gamma$ on $\pb$ by projecting all off-shell
momenta radially onto the Fermi surface \cite{RevModPhys.84.299}.
Then, $\Gamma$ is
reduced  to
$\Gamma_{1',2';1,2}\rightarrow\Gamma(\pb_{F1}',\pb_{F2}',\pb_{F1})$ where
the last momentum variable is dropped because it is fixed by
conservation, and the
subscript in $\pb_F$ indicates radial projection onto the Fermi surface.
The initial condition for $\Gamma$ at the ultraviolet scale $\Lambda_{UV}$
is given by the antisymmetrized bare interaction $V(\qb)$,
\begin{equation}
  \Gamma(\pb_{F1}',\pb_{F2}',\pb_{F1})\big|_{\Lambda_{UV}}
  \equiv \frac{1}{2}
  [V(\pb_{F1}'-\pb_{F1})-V(\pb_{F2}'-\pb_{F1})].
\end{equation}
We solve the flow equation by the Euler method on a logarithmic grid of
$\Lambda$ consisting of  $10^3$ RG steps going from $\Lambda_{UV}=0.99E_F$
down to $\Lambda_{IR}=10^{-3}E_F$.  Each $\pb_{F}$ is discretized on an angular
grid with up to hundreds of patches on the Fermi surface \footnote{To speed up
  the calculation, the FRG algorithm is adapted to run parallel on Graphic
  Processing Units.}. 
We monitor the flow of $\Gamma(\pb_{F1}',\pb_{F2}',\pb_{F1})$
which contains hundreds of millions of running coupling constants. 
When the absolute value of a running coupling constant in $\Gamma$ exceeds a
threshold, e.g. $50E_F$, signaling an imminent divergence, we terminate the
flow, record the critical scale $\Lambda_c$, and analyze the vertex to
diagnose the instability.
If the flow continues smoothly down to 
$\Lambda_{IR}$, we conclude the Fermi liquid is stable down to
exponentially small temperatures.
Scanning the parameter space $(g,r_c)$ gives the phase diagram, whereas
$\Lambda_c$ provides a rough estimate of the $T_c$ of each ordered phase.

Two complementary methods are employed to identify the leading
instability from the large, complex data set of $\Gamma$.
First, we plot $\Gamma(\pb_{F1}',\pb_{F2}',\pb_{F1})$ at $\Lambda_c$ against
the angular directions of $\pb_{F1}'$ and $\pb_{F2}'$ for fixed
$\pb_{F1}=(-p_F,0)$ \footnote{This is done without loss of generality due to
  the rotational invariance.} to reveal the dominant correlations between
particles on the Fermi surface. The color map (Fig.~\ref{fig:phase_diagram},
lower right columns) shows
diagonal structures ($\pb_{F1}'=-\pb_{F2}'$) for pairing instability, and
horizontal-vertical structures (scattering $\pb_{F1}\rightarrow \pb_{F1}'$
with momentum transfer close to $0$ or $2p_F$) for density waves
\cite{PhysRevB.63.035109,peterkopietz2010}.  This method directly exposes the
pairing symmetry through the number of nodes along the diagonal structures: a
$p$-wave phase has one node, and an $f$-wave phase has three nodes, etc.
In the second method, we construct the channel matrices from $\Gamma$, e.g.
$V_{BCS}(\pb',\pb)=\Gamma(\pb',-\pb',\pb)$ for the pairing channel, and
$V^\qb_{DW}(\pb',\pb)=\Gamma(\pb+\qb/2,\pb'-\qb/2,\pb-\qb/2)$ for the density
wave channel. Different values of $\qb$, e.g. $\qb_i=(q_i,0)$ with
$q_i\in\{0.05p_F,0.5p_F,p_F,2p_F\}$ for $i\in\{1,...,4\}$ respectively, are
compared (see DW$_i$ in Fig.~\ref{fig:phase_diagram},
left column).
The channel matrices are then diagonalized and their the most negative
eigenvalues are monitored.  This method provides a clear picture of the
competition among the channels. The eigenvector of the
leading divergence exposes the orbital symmetry, e.g. $p$- or $f$-wave, of the
incipient order. 


\section{Phase diagram from FRG}
\label{sec:results}
The resulting phase diagram is summarized in
the top panel of Fig.~\ref{fig:phase_diagram}. In addition to the Fermi
liquid, three ordered phases are clearly identified. 
Here the filled circles mark the phase boundary,  the color
indicates the critical scales $\Lambda_c$ which is proportional to $T_c$ \cite{RevModPhys.84.299},
and the dash lines are guide for the eye and they roughly enclose the regions where 
$\Lambda_c$ is higher than the numerical IR scale $\Lambda_{IR}$. 
For attractive interactions $g<0$, e.g. at the point $\mathcal{P}_1$, the
leading eigenvalues are from $V_{BCS}$ and doubly degenerate  with $p$-wave
symmetry. 
The vertex map also reveals
diagonal structures with single node (Fig.~\ref{fig:phase_diagram}),
confirming a $p$-wave
superfluid phase.  While the FRG here cannot directly access the wavefunction
of the broken symmetry phase, mean field argument favors a $p_x+ip_y$ ground
state because it is fully gapped and has the most condensation energy.  Thus
Rydberg-dressed Fermi gas is a promising system to realize the $p_x+ip_y$
topological superfluid.  Our analysis suggests that the optimal $T_c$ is
around $p_Fr_c\sim 2$ and $T_c$ increases with $|u_0|$. 

For repulsive interactions $g>0$, which channel gives the leading instability
depends intricately on the competition between $p_F$ and $r_c$.
{\bf (a)} First, FRG reveals a density wave phase for $p_Fr_c \gtrsim 4$, in
broad agreement with RPA.  For example, at point $\mathcal{P}_3$, the most
diverging eigenvalue comes from $V_{DW}$, and the vertex map shows clear
horizontal-vertical structures (Fig.~\ref{fig:phase_diagram}). Note the
separations between the
horizontal/vertical lines, and relatedly the ordering wave vector, depend on
$r_c$.
{\bf (b)} For $p_Fr_c\lesssim 4$, however, the dominant instability comes from
the BCS channel despite that the bare interaction is purely repulsive in real
space.  In particular, for small $p_Fr_c\lesssim2$, such as the point
$\mathcal{P}_2$ in Fig.~\ref{fig:phase_diagram}, the pairing symmetry can be
unambiguously
identified to be $f$-wave: the vertex map has three nodes, the most diverging
eigenvalues of $V_{BCS}$ are doubly degenerate, and their eigenvectors follow
the form $e^{\pm i 3\theta}$.  
This $f$-wave superfluid is the most striking
result from FRG.  {\bf (c)} 
For $p_Fr_c$ roughly between 2 and 4, sandwiched
between the density wave and $f$-wave superfluid, lies a region 
where the superfluid paring channel strongly intertwines with the density wave channel. 
While the leading divergence is still superfluid, it is no longer pure
$f$-wave, and it becomes increasingly degenerate with a subleading density wave order.  
This hints at a coexistence of superfluid and density wave.

To determine the phase boundary, we trace the evolution of $\Lambda_c$ along a few vertical cuts in
the phase diagram, and use the kinks in $\Lambda_c$ as indications for the
transition between the density wave and superfluid phase, or a change in
pairing symmetry within the superfluid  (see inset, top panel of
Fig.~\ref{fig:phase_diagram}). We
have checked the phase boundary (filled circles) determined this way  is
consistent with the eigenvalue flow and vertex map.  

Cooper pairing can occur in repulsive Fermi liquids via the Kohn-Luttinger
(KL) mechanism through  the
renormalization of fermion vertex by the particle-hole fluctuations. Even for featureless bare interactions $V(\qb)=U>0$, the
effective interaction $V_{\ell}$ in angular momentum channel $\ell$ can become
attractive due to over-screening  by the remaining fermions
\cite{PhysRevLett.15.524}. 
In 2D, 
the KL mechanism becomes effective at
higher orders of perturbation theory, e.g. $U^3$, and the leading pairing
channel is believed to be $p$-wave \cite{PhysRevB.48.1097}. Here,  the
effective interaction is also
strongly renormalized from the bare interaction by 
particle-hole fluctuations. 
We have checked that turning off the ZS and ZS' channels eliminates
superfluid order on the repulsive side. 
%
%
However, our system exhibits $f$-wave pairing with a significant critical
temperature in contrast to usual KL mechanism with exponentially small $T_c$.
%
%
This is because the Rydberg-dressed
interaction already contains a ``pairing seed": $V(\qb)$ develops
a negative minimum in momentum space for $q=q_c$ unlike the featureless
interaction $U$. 
Among all the scattering processes $(\pb_{F},-\pb_{F})\rightarrow
(\pb'_F,-\pb'_F)$, those with $q=|\pb'_F-\pb_F|\sim q_c$ favor pairing.  It
follows that pairing on the repulsive side occurs most likely when
the Fermi surface has a proper size, roughly $2p_F\sim q_c$, in broad
agreement with the FRG phase diagram. These considerations based on the bare
interaction and BCS approach, however, are insufficient to explain the
$f$-wave superfluid revealed only by FRG, which accurately accounts the
interference between the particle-particle and particle-hole channels. 
The pairing seed and over screening conspire to give
rise to a robust $f$-wave superfluid with significant $T_c$. 


\section{Conclusion}
\label{sec:conclusions}

We developed an unbiased numerical 
technique based on FRG to obtain the phase diagram for the new system of Rydberg-dressed Fermi gas  to guide future experiment.
We found an $f$-wave superfluid with unexpectedly high $T_c$ driven by repulsive interactions beyond the conventional Kohn-Luttinger paradigm. The physical mechanism behind the $T_c$ enhancement is traced back to the negative minimum in the bare interaction, as well as the renormalization
of the effective interaction by particle-hole fluctuations. These results contribute to our understanding of unconventional pairing from repulsive interactions, and more generally, competing many-body instabilities of fermions with long-range interactions. 
Our analysis may be used for optimizing $T_c$ by engineering effective interactions using schemes similar to Rydberg dressing.
Our FRG approach can also be applied to illuminate the rich interplay of competing density wave and pairing fluctuations in solid state 
correlated quantum materials. Note that f-wave
pairing has been previously discussed in the context of fermions on the
p-orbital bands \cite{PhysRevA.82.053611, PhysRevB.83.144506}.

\begin{acknowledgments}
This work is supported by 
NSF Grant No. PHY-1707484, 
AFOSR Grant No.  FA9550-16-1-0006 (A.K. and E.Z.),
ARO Grant No. W911NF-11-1-0230, and 
MURI-ARO Grant No. W911NF-17-1-0323 (A.K.). 
X.L. acknowledges support by National Program on Key Basic Research Project of China under Grant No. 2017YFA0304204 and National Natural Science Foundation of China under Grants No. 11774067.
\end{acknowledgments}
\bibliography{refs}
\end{document}